\documentclass{llncs}
\usepackage{amsmath,amstext,amssymb}

\title{Choiceless Polynomial Space\thanks{The work of the first author has been funded by the Federal Ministry for Climate Action, Environment, Energy, Mobility, Innovation and Technology (BMK), the Federal Ministry for Digital and Economic Affairs (BMDW), and the State of Upper Austria in the frame of the COMET Module Dependable Production Environments with Software Security (DEPS) within the COMET - Competence Centers for Excellent Technologies Programme managed by Austrian Research Promotion Agency FFG.}}

\author{Flavio Ferrarotti\inst{1}, Klaus-Dieter Schewe\inst{2}}

\institute{Software Competence Centre Hagenberg, Hagenberg, Austria, \email{flavio.ferrarotti@scch.at},
\and
Linz, Austria, \email{kd.schewe@liwest.at}}

\begin{document}

\maketitle
\thispagestyle{plain}

\begin{abstract}

Abstract State Machines (ASMs) provide a model of computations on structures rather than strings. Blass, Gurevich and Shelah showed that deterministic PTIME-bounded ASMs define the choiceless fragment of PTIME, but cannot capture PTIME. In this article deterministic PSPACE-bounded ASMs are introduced, and it is proven that they cannot capture PSPACE. The key for the proof is a characterisation by partial fixed-point formulae over the St\"ark/Nanchen logic for deterministic ASMs and a construction of transitive structures, in which such formulae must hold\footnote{This construction exploits that the decisive support theorem for choiceless polynomial time holds under slightly weaker assumptions. To that extent our work is mainly grounded in this observation.}.


\end{abstract}

\section{Introduction}\label{sec:intro}

Abstract State Machines (ASMs) provide a model of computations on structures, which serves both for research on theory of computation and for rigorous development of software systems \cite{boerger:2003}. Reasoning about ASMs is further supported by the St\"ark/Nanchen logic for deterministic ASMs \cite{staerk:jucs2001}, which has been extended to a logic for non-deterministic ASMs by Ferrarotti et al. \cite{ferrarotti:amai2018}. The behavioural theories of ASMs cover the capture of sequential and recursive algorithms \cite{gurevich:tocl2000,boerger:fi2020} as well as synchronous and asynchronous parallel algorithms \cite{ferrarotti:tcs2016,boerger:ai2016} on arbitrary levels of abstraction. Thus, ASMs provide a promising candidate for a generalised theory of computation on structures.

Concerning complexity Blass, Gurevich and Shelah investigated the choiceless fragment of PTIME defining {\em Choiceless Polynomial Time} (CPT) \cite{blass:apal1999}. They showed that CPT is a PTIME logic in the sense of Gurevich \cite{gurevich:cttcs1988}, which does not capture PTIME. Gurevich even conjectured that no logic capturing PTIME exists.

In this paper we approach an analogous characterisation of the choiceless fragment of PSPACE by defining {\em Choiceless Polynomial Space} (CPS) and proving that PSPACE is not captured by this logic. However, as PSPACE = NPSPACE holds, this complexity class is easily captured by exploiting non-deterministic ASMs.

While we assume familiarity with ASMs and basic complexity theory, we start with a few essential preliminaries in Section \ref{sec:prel} comprising ASMs with base sets defined by hereditarily finite sets, which lead in a straightforward way to the model of deterministic PSPACE-bounded ASMs. We continue in Section \ref{sec:cps} defining our model of CPS. Then by exploiting the logic of deterministic ASMs we obtain a partial fixed-point formula characterising a PSPACE computation, which must hold in a transitive structure containing all active objects. This is quite analogous to the work on CPT; we have to deal with partial fixed-points instead of inflationary ones.

Furthermore, we obtain a Support Theorem in Section \ref{sec:support} for the case of an empty input signature. Its proof was already contained in the proof of the Support Theorem in \cite{blass:apal1999}, which still holds under weaker assumptions. As the partial fixed-point logic FO[PFP] can be embedded into the infinitary logic $\mathcal{L}^{\omega}_{\omega\infty}$, we can exploit the Equivalence Theorem from \cite{blass:apal1999} for sufficiently large structures. Then we need a winning strategy of the duplicator for the Ehrenfeucht-Fra\"{\i}ss\'{e} pebble game characterising equivalent structures under $\mathcal{L}^{\omega}_{\omega\infty}$. Using the Equivalence Theorem we obtain immediately that Parity is not separable by CPS, hence CPS cannot capture PSPACE; it does not even subsume all of PTIME. 

\section{Preliminaries}\label{sec:prel}

ASMs including their foundations, semantics and usage in applications are the subject of the detailed monograph by B\"orger and St\"ark \cite{boerger:2003}. In a nutshell, an ASM is defined by a signature, i.e. a finite set of function (and relation) symbols, a background, and a rule. The signature defines states as structures, out of which a set of initial states is defined. The sets of states and initial states are closed under isomorphisms. The background defines domains and fixed operations on them that appear in every state \cite{blass:beatcs2007}, and the rule defines a function mapping states to successor states. Following \cite{blass:apal1999} we use base sets defined by hereditarily finite sets.

\subsection{States}

The {\em background} of an ASM, as we use them here, comprises logic names and set-theoretic names:

\begin{description}

\item[Logic names] comprise the binary equality $=$, nullary function names \textbf{true} and \textbf{false} and the usual Boolean operations. All logic names are relational.

\item[Set-theoretic names] comprise the binary predicate $\in$, nullary function names $\emptyset$ and \textit{Atoms\/}, unary function names $\bigcup$ and \textit{TheUnique\/}, and the binary function name \textit{Pair\/}. 

\end{description}

As in \cite{blass:apal1999} we will use $\emptyset$ also to denote undefinedness, for which usually another function name \textit{undef\/} would be used. In this way we can concentrate on sets.

The signature $\Upsilon$ of an ASM, as we use them here, comprises input names and dynamic names:

\begin{description}

\item[Input names] are given by a finite set of relation symbols, each with a fixed arity. Input names will be considered being static, i.e. locations defined by them will never be updated by the ASM.

\item[Dynamic names] are given by a finite set of function symbols, each with a fixed arity, including \textit{Output\/} and a nullary function symbol \textit{Halt\/}. Some of the dynamic names may be relational. We use the notation $\Upsilon_{\text{dyn}}$ for the set of dynamic function symbols.

\end{description}

{\em States} are defined as structures over the signature $\Upsilon$ plus the background signature, for which we assume specific base sets. A {\em base set} $B$ comprises two parts: a finite set $A$ of {\em atoms}, which are not sets, and the collection $B = \textit{HF\/}(A)$ of hereditarily finite sets built over $A$. The set $\textit{HF}(A)$ is the smallest set such that if $x_1, \ldots, x_n$ are in $A \cup \mathit{HF}(A)$, then $\{x_1, \ldots, x_n\}$ is in $\mathit{HF}(A)$. Each element $x \in \textit{HF\/}(A)$ has a well-defined {\em rank} $rk(x)$. We have $rk(x) = 0$, if $x = \emptyset$ or $x$ is an atom. If $x$ is a non-empty set, we define its rank as the smallest ordinal $\alpha$ such that $rk(y) < \alpha$ holds for all $y \in x$. Then the atoms in $A$ and the sets in $\textit{HF\/}(A)$ are called the {\em objects} of the base set $B = \textit{HF\/}(A)$. A set $X$ is called {\em transitive} iff $x \in X$ and $y \in x$ implies $y \in X$. If $x$ is an object, then $\textit{TC\/}(x)$ denotes the least transitive set $X$ with $x \in X$. If $\textit{TC\/}(x)$ is finite, the object $x$ is called {\em hereditarily finite}. In this work we often talk of {\em transitive structures}, meaning structures with transitive base sets.

The logic names are interpreted in the usual way, i.e. \textbf{true} and \textbf{false} are interpreted by 1 and 0, respectively (i.e. by $\{ \emptyset \}$ and $\emptyset$). Boolean operations are undefined, i.e. give rise to the value 0, if at least one of the arguments is not Boolean. An {\em isomorphism} is a permutation $\sigma$ of the set $A$ of atoms that is extended to sets in $B$ by $\sigma(\{ b_1, \dots, b_k \}) = \{ \sigma(b_1), \dots, \sigma(b_k) \}$.

The set-theoretic names $\in$ and $\emptyset$ are interpreted in the obvious way, and \textit{Atoms\/} is interpreted by the set of atoms of the base set. If $a_1 ,\dots, a_k$ are atoms and $b_1 ,\dots, b_\ell$ are sets, then $\bigcup \{ a_1 ,\dots, a_k, b_1 ,\dots, b_\ell \} = b_1 \cup\dots\cup b_\ell$. For $b = \{ a \}$ we have $\textit{TheUnique\/}(b) = a$, otherwise it is undefined. Furthermore, we have $\textit{Pair\/}(a,b) = \{ a, b \}$.

An input name $p$ is interpreted by a Boolean-valued function. If the arity is $n$ and $p(a_1, \dots, a_n)$ holds, then each $a_i$ must be an atom. Finally, a dynamic function symbol $f$ of arity $n$ is interpreted by a function $f_S: B^n \rightarrow B$ (or by $f_S: B^n \rightarrow \{ 0, 1 \}$, if $f$ is relational). The domain $\{ (a_1 , \dots, a_n) \mid f(a_1 ,\dots, a_n) \neq 0 \}$ is required to be finite. With such an interpretation we obtain the set of states over the signature $\Upsilon$ and the given background. 

An {\em input structure} is a finite structure $I$ over the subsignature comprising only the input names.  We assumed that only atoms appear in $I$. If the finite set of atoms in the input structure is $A$, then $|A|$ is referred to as the {\em size} of the input. An {\em initial state} $S_0$ is a state over the base set $B = \textit{HF\/}(A)$ which extends $I$ such that the domain of each dynamic function is empty. We call $S_0 = \text{State}(I)$ the {\em initial state generated by $I$}. To emphasise the dependence on $I$, we also write $\textit{HF\/}(I)$ instead of $B$.

\subsection{Terms and Rules}

{\em Terms} are defined in the usual way from a set of variables $V$, function symbols in the signature $\Upsilon$ and the set constructor $\{t(v) \mid v \in s \wedge g(v)\}$, where $v$ is a variable that does not occur free in term $s$. The semantics of $\{t(v) \mid v \in s \wedge g(v)\}$ is the obvious for set comprehension. That is, the term language is the same as for CPT \cite{blass:apal1999}. Then the set $\textit{fr\/}(t)$ of {\em free variables} in a term $t$ is again defined as usual. In particular,  for set comprehension terms we have $\textit{fr\/}(\{ t(v) \mid v \in s \wedge g(v) \}) = (\textit{fr\/}(t(v)) \cup \textit{fr\/}(s) \cup \textit{fr\/}(g(v))) - \{ v \}$. Also the interpretation of terms in a state $S$ is standard.

{\em ASM rules} are defined as follows:

\begin{description}

\item[skip] is a rule.

\item[assignment.] If $f$ is a dynamic function symbol in $\Upsilon$ of arity $n$ and $t_0, \dots, t_n$ are terms, then $f(t_1, \dots, t_n) := t_0$ is a rule.

\item[branching.] If $\varphi$ is a Boolean term and $r_1$, $r_2$ are rules, then also \textbf{if} $\varphi$ \textbf{then} $r_1$ \textbf{else} $r_2$ \textbf{endif} is a rule. 
We also use the shortcut \textbf{if} $\varphi$ \textbf{then} $r_1$ \textbf{endif} for \textbf{if} $\varphi$ \textbf{then} $r_1$ \textbf{else} \textbf{skip} \textbf{endif}.

\item[parallelism.] If $v$ is a variable, $t$ is a term with $v \notin \textit{fr\/}(t)$, and $r(v)$ is a rule, then also \textbf{forall} $v \in t$ \textbf{do} $r(v)$ \textbf{enddo} is a rule.
We also use the shortcut \textbf{par} $r_1 \dots r_k$ \textbf{endpar} for \textbf{forall} $i \in \{ 1,\dots,k \}$ \textbf{do} \textbf{if} $i=1$ \textbf{then} $r_1$ \textbf{else} \textbf{if} $i=2$ \textbf{then} $r_2$ \textbf{else} \dots \textbf{if} $i=k$ \textbf{then} $r_k$ \textbf{endif} \dots \textbf{endif} \textbf{enddo}.

\end{description}

The rule associated with an ASM must be closed, i.e., it must not have free variables. If $f$ is dynamic function symbol in $\Upsilon$ of arity $n$, and $a_1, \dots, a_n$ are objects of the base set $B$ of a state $S$, then the pair $(f,(a_1, \dots, a_n))$ is a {\em location} of the state $S$. We use the abbreviation $\bar{a}$ for tuples $(a_1, \dots, a_n)$, whenever the arity is known from the context. For a location $\ell = (f,\bar{a})$ we write $val_S(\ell) = b$ iff $f_S(a_1, \dots, a_n) = b$; we call $b$ the value of the location $\ell$ in the state $S$. An {\em update} is a pair $(\ell, a)$ consisting of a location $\ell$ and an object $a \in B$, and an {\em update set} (for a state $S$) is a set of updates with locations of $S$ and objects $a$ in the base set of $S$.

Now let $S$ be a state with base set $B$, and let $\zeta : V \rightarrow B$ be a variable assignment. Let $r$ be an ASM rule. We define an update set $\Delta_{r,\zeta}(S)$ on state $S$ for the rule $r$ depending on $\zeta$ as follows:

\begin{itemize}

\item $\Delta_{\textbf{skip},\zeta}(S) = \emptyset$.

\item For an assignment rule $r$ of the form $f(t_1 ,\dots, t_n) := t_0$ we have $\Delta_{r,\zeta}(S) = \{ (\ell,a) \}$ with the location $\ell = (f, (\text{val}_{S,\zeta}(t_1) ,\dots, \text{val}_{S,\zeta}(t_n)))$ and the object $a = \text{val}_{S,\zeta}(t_0)$.

\item For a branching rule $r$ of the form \textbf{if} $\varphi$ \textbf{then} $r_1$ \textbf{else} $r_2$ \textbf{endif} we have $\Delta_{r,\zeta}(S) = \Delta_{r_1,\zeta}(S)$, if $\text{val}_{S,\zeta}(\varphi) = 1$ holds, and $\Delta_{r,\zeta}(S) = \Delta_{r_2,\zeta}(S)$ else.

\item For a parallel rule $r$ of the form \textbf{forall} $v \in t$ \textbf{do} $r(v)$ \textbf{enddo} we have
$\Delta_{r,\zeta}(S) = \bigcup_{a \in \text{val}_{S,\zeta}(t)}$ $\Delta_{r(v),\zeta(v \mapsto a)}(S)$.

\end{itemize}

\subsection{PSPACE ASMs}

An update set $\Delta$ is {\em consistent} iff for any two updates $(\ell,a_1), (\ell,a_2) \in \Delta$ with the same location we have $a_1 = a_2$. This defines the notion of {\em successor state} $S^\prime = S + \Delta$ of a state $S$. For a consistent update set $\Delta = \Delta_{r,\zeta}(S)$ and a location $\ell$ we have $\text{val}_{S^\prime}(\ell) = a$ for $(\ell,a) \in \Delta$, and $\text{val}_{S^\prime}(\ell) = \text{val}_S(\ell)$ else. In addition, let $S + \Delta = S$ for inconsistent update sets $\Delta$.

A {\em run} of an ASM $M$ with rule $r$ is a finite or infinite sequence of states $S_0, S_1, \dots$ such that $S_0$ is an initial state and $S_{i+1} = S_i + \Delta_r(S_i)$ holds. Furthermore, if $k$ is the length of a run ($k = \omega$ for an infinite run), then \textit{Halt\/} must fail on all states $S_i$ with $i < k$. Note that in a run all states have the same base set, which is in accordance with requirements from the behavioural theories of sequential and parallel algorithms \cite{gurevich:tocl2000,ferrarotti:tcs2016}.

An object $a \in B$ is called {\em critical} in state $S$ iff $a$ is an atom or $a \in \{ 0, 1 \}$ or $a$ is the value of a location $\ell$ of $S$ or there is a location $\ell = (f, \bar{a})$ with  $\text{val}_S(\ell) \neq \emptyset$ and $a$ appears in $\bar{a}$. An object $a \in B$ is called {\em active} in $S$ iff there exists a critical object $a^\prime$ with $a \in \textit{TC\/}(a^\prime)$. In addition, if $\rho = S_0, S_1, \dots$ is a run of an ASM, then we call an object $a \in B$ {\em active} in $\rho$ iff $a$ is active in at least one state $S_i$ of $\rho$.

We define PSPACE(-bounded) ASMs by requesting a polynomial bound on the number of objects that can be active in any state of its runs. This implies that in every state in a computation of a PSPACE ASM the number of locations in use (i.e. those that have a value different than $0$), as well as the size of the objects stored in these locations, will also have a polynomial bound.

A {\em PSPACE ASM} is a pair $\tilde{M} = (M, p(n))$ comprising an ASM $M$ and an integer polynomial $p(n)$. A \emph{run} of $\tilde{M}$ with initial state $S_0$ generated by an input structure $I$ of size $n$ is the longest initial segment $\rho$ of the run of $M$ on $S_0$ such that for each $S \in \rho$ the number of active objects in $S$ is at most $p(n)$.

A PSPACE ASM $\tilde{M}$ \emph{accepts} an input structure $I$ iff the run of $\tilde{M}$ with initial state $\text{State}(I)$ ends in a state with value $1$ for $\mathit{Halt}$ (i.e. it is finite), and the value of $\mathit{Output}$ is $1$. Analogously, $\tilde{M}$ \emph{rejects} $I$ iff the run of $\tilde{M}$ with initial state $\text{State}(I)$ ends in a state with value $1$ for $\mathit{Halt}$ and the value of $\mathit{Output}$ is $0$.

\section{Choiceless Polynomial Space}\label{sec:cps}

The complexity class {\em Choiceless Polynomial Space} (CPS) is the collection of pairs $(K_1, K_2)$, where $K_1$ and $K_2$ are disjoint classes of finite structures of the same signature, such that there exists a PSPACE ASM that accepts all structures in $K_1$ and rejects all structures in $K_2$.

We also say that a pair $(K_1, K_2) \in$ CPS is {\em CPS separable}. As for the analogous definition of CPT a PSPACE ASM may accept structures not in $K_1$ and reject structures not in $K_2$. Therefore, we also say that a class $K$ of finite structures is in CPS, if $(K,K^\prime) \in$ CPS holds for the complement $K^\prime$ of structures over the same signature.

According to Gurevich \cite{gurevich:cttcs1988} a logic $\mathcal{L}$ can be defined in general by a pair (\textit{Sen\/},\textit{Sat\/}) of functions satisfying the following conditions:

\begin{itemize}

\item \textit{Sen\/} assigns to every signature $\Upsilon$ a recursive set $\textit{Sen\/}(\Upsilon)$, the set of {\em $\mathcal{L}$-sentences of signature $\Upsilon$}.

\item \textit{Sat\/} assigns to every signature $\Upsilon$ a recursive binary relation $\textit{Sat\/}_\Upsilon$ over structures $S$ over $\Upsilon$ and sentences $\varphi \in \textit{Sen\/}(\Upsilon)$. We assume that whenever $S$ and $S^\prime$ are isomorphic, then $\textit{Sat\/}_\Upsilon(S,\varphi) \Leftrightarrow \textit{Sat\/}_\Upsilon(S^\prime,\varphi)$ holds.

\end{itemize}

We say that a structure $S$ over $\Upsilon$ {\em satisfies} $\varphi \in \textit{Sen\/}(\Upsilon)$ (notation: $S \models \varphi$) iff $\textit{Sat\/}_\Upsilon(S,\varphi)$ holds. 
If $\mathcal{L}$ is a logic in this general sense, then for each signature $\Upsilon$ and each sentence $\varphi \in \textit{Sen\/}(\Upsilon)$ let $K(\Upsilon,\varphi)$ be the class of structures $S$ of signature $\Upsilon$ with $S \models \varphi$. We then say that $\mathcal{L}$ is a {\em PSPACE logic}, if every class $K(\Upsilon,\varphi)$ is PSPACE in the sense that it is closed under isomorphisms and there exists a PSPACE Turing machine that accepts exactly the standard encodings of ordered versions of the structures in the class.

We further say that a logic $\mathcal{L}$ {\em captures} PSPACE iff it is a PSPACE logic and for every signature $\Upsilon$ every PSPACE class of $\Upsilon$-structures coincides with some class $K(\Upsilon,\varphi)$ and $\varphi \in \textit{Sen\/}(\Upsilon)$.

These definitions of PSPACE logics can be generalised to three-valued logics, in which case $\textit{Sat\/}_\Upsilon(S,\varphi)$ may be true, false or unknown. For these possibilities we say that $\varphi$ {\em accepts} $S$ or $\varphi$ {\em rejects} $S$ or neither, respectively. Then two disjoint classes $K_1$ and $K_2$ of structures over $\Upsilon$ are called {\em $\mathcal{L}$-separable} iff there exists a sentence $\varphi$ accepting all structures in $K_1$ and rejecting all those in $K_2$.

In this sense, CPS defines a three-valued PSPACE logic that separates pairs of structures in CPS. The idea is that sentences of this logic are PSPACE ASMs, for which $\Upsilon$ is the signature of the input structure. By abuse of terminology we also denote this logic as CPS.

Let $\tilde{M} = (M, p(n))$ be a PSPACE ASM, and let $\mathit{Active}(I)$ denote the set of active objects in the run of $\tilde{M}$ on $\mathit{State}(I)$. Note that due to the definition of active objects, this set is transitive. By abuse of notation, let $\mathit{Active}(I)$ also denote the structure $(\mathit{Active(I)}, \bar{R})$ plus background structure, where $\bar{R}$ stands for all the relations in the input structure $I$. Let $\rho$ be a run of $\tilde{M}$ on $\mathit{State}(I)$. 
Notice that $\rho$ could be \emph{infinite}, as the run of $M$ on $\mathit{State}(I)$ can be infinite and never violate the defining conditions of PSPACE ASMs.
For each $f \in \Upsilon_{\text{dyn}}$ we introduce a new relation symbol $D_f$ with the intended interpretation that $D_f(\bar{x}, y)$ should hold iff the run $\rho$ is finite and $S_l \models f(\bar{x}) = y \neq \emptyset$, where $S_l$ is the final state in $\rho$. 

Analogous to \cite[Thm.~18]{blass:apal1999} we obtain a Fixed-Point-Theorem. Before formulating this theorem and its proof, let us observe that w.l.o.g. we can write every formulae of first-order logic in an equivalent term normal form, where all atomic subformulae that are equations with a function symbol $g$ take the form $g(\bar{x}) = t$, where $\bar{x}$ is a tuple of variables and $t$ is either a variable or a constant \textbf{true} or \textbf{false}. Then we can exploit the St\"ark/Nanchen logic for ASMs \cite{staerk:jucs2001} (see also \cite[Sect.~8.1]{boerger:2003}), which is a definitional extension of first-order logic. Most decisively, we obtain a first-order formula $upd_{r,f}(\bar{x}, y)$ such that for all states $S$ and all variable assignments $\zeta$ we have that $S, \zeta \models upd_{r,f}(\bar{x}, y)$ iff $(f, \bar{x}, y) \in \Delta_{r, \zeta}(S)$ and $\Delta_{r, \zeta}(S)$ is consistent.  

\begin{theorem}[Fixed-Point Theorem]\label{thm:pfp}
The relations $D_f(\bar{x}, y)$ for $f \in \Upsilon_{\text{dyn}}$ are uniformly definable (i.e. independently of the input structure $I$) on $\mathit{Active}(I)$ by a partial fixed-point formula.

\end{theorem}

\begin{proof}

Let $r$ be the rule of the ASM $\tilde{M}$ and let $\rho$ be its run. The relation $D_f$, where $f$ ranges over $\Upsilon_{\text{dyn}}$, is the partial fixed-point defined by simultaneous induction on $\mathit{Active}(I)$ using the rule
\[ D_f(\bar{x}, y) = y \neq 0 \wedge \big(U_{r,f}(\bar{x}, y) \vee \big(D_f(\bar{x},y) \wedge \neg \exists z (z \neq y \wedge U_{r,f}(\bar{x}, z))\big)\big) \; ,\]
where $U_{r,f}(\bar{x}, y)$ is obtained by replacing in $upd_{r,f}(\bar{x}, y)$ every atomic subformula of the form $g(\bar{t}) = s$ with dynamic function symbol $g$ by $D_g(\bar{t}, s)$. Note that if $S_i$ is the $i$-th state in $\rho$, then at stage $i$ of the simultaneous induction defined by this rules we get that $D_f(\bar{x}, y)$ iff $S_i \models f(\bar{x}) = y \neq \emptyset$. If a run $\rho = S_0, \ldots, S_l$ is finite, then a fixed-point will be reached after $l$ steps and $D_f(\bar{x}, y)$ will hold iff $S_l \models f(\bar{x}) = y \neq \emptyset$. Otherwise, if $\rho$ is infinite, then the simultaneous induction will not reach a fixed-point and $D_f$ will be the empty relation.\qed

\end{proof}

Then $D_{\textit{Halt\/}}(1) \wedge D_{\textit{Output\/}}(1)$ expresses that the ASM will terminate and produce the output \textbf{true}.

\section{Limitations of CPS}\label{sec:support}

Theorem \ref{thm:pfp} states that a problem is in CPS, if we find a formula in the partial fixed-point logic FO[PFP] (see \cite{ebbinghaus:1995}) that holds in a transitive structure (or equivalently in all) transitive structures that contain all active objects of the CPS computation corresponding to this formula iff the input structure is accepted. In this we want to show that there are PSPACE problems not in CPS. For this we only consider CPS computations where the input signature is empty, i.e. the input structure $I$ degenerates to a naked set. 

\subsection{A Support Theorem}

We first investigate suitable transitive structures that will contain all active objects without having to consider specific PSPACE ASMs. Let $Aut(I)$ be the automorphism group of the input structure $I$, which for naked sets is the group of all permutations of the atoms. Every automorphism extends naturally to $State(I)$. A {\em support set} of an object $y$ is a set $S$ of atoms such that every automorphism $\pi$ with $\pi(x) = x$ for all $x \in S$ also satisfies $\pi(y) = y$.

Our first aim is to show the following Support Theorem using the constant $k$ determined by the polynomial space bound, i.e. $| Active(S) | \le n^k$ holds, where $Active(S)$ is the set of active objects in state $S$ and $n$ is the size of the set of atoms. The proof is the same as the proof of the Support Theorem for CPT \cite[Thm.~24]{blass:apal1999}. 

\begin{theorem}[Support Theorem]\label{thm-support}
If $n = | I |$ is sufficiently large, then every active object $y$ has a unique minimal support set $\text{Supp}(y)$ of cardinality $| \text{Supp}(y) | \le k$.
\end{theorem}

First we notice that if $X_1$, $X_2$ are support sets of an object $y$ with $X_1 \cup X_2 \neq I$, then also $X_1 \cap X_2$ supports $y$. This is Lemma 26 in \cite{blass:apal1999}, and the proof holds without any change. Consequently, if there exists a support set $X$ of size $< n/2$, then there exists a unique minimal support set $Supp(y) = \bigcap \{ X \mid X \text{ supports } y \text{ and } |X| < n/2 \}$.

\begin{lemma}\label{lem-support2}

Assume that $n$ is large enough such that $\dbinom{n}{k + 1} > n^k$ holds. If an active object $y$ has a support set $X$ with $|X| < n/2$, then $| \text{Supp}(y) | \le k$ holds.

\end{lemma}

\begin{proof}

Suppose $y$ has a support set $Supp(y)$ of size $s < n/2$. Any automorphism $\pi$ with $\pi(y) = z$ satisfies $\pi(Supp(y)) = Supp(z)$. Assume that $s > k$ holds. Then for large enough $n$ we get the contradiction
\begin{align*}
n^k \ge & | Active(S) | \ge | \{ \pi(y) \mid \pi \in Aut(I) \} | \ge \\
& | \{ \pi(Supp(y)) \mid \pi \in Aut(I) \} | = \dbinom{n}{s} \ge \dbinom{n}{k + 1} > n^k
\end{align*}
Hence $s \le k$ holds.\qed

\end{proof}

\begin{lemma}\label{lem-support3}

If $n = | I |$ is sufficiently large, then every active object $y$ has a support $X$ with $|X | < n/2$. 

\end{lemma}

The proof is the same as the proof of Lemma 28 in \cite{blass:apal1999} without any change. Lemmata \ref{lem-support2} and \ref{lem-support3} together imply the Support Theorem \ref{thm-support}.

\subsection{Symmetric Objects}

With the Support Theorem \ref{thm-support} the results in \cite[Section~9]{blass:apal1999} remain valid for CPS. We call an object $y \in \textit{HF\/}(I)$ {\em $k$-symmetric} for some positive integer $k$ iff every $z \in TC(y)$ has a support set of size $\le k$. Concentrate on the special case, where the input signature $\Upsilon_0$ is empty, thus input structures are simply naked sets. Then let $I_k$ denote the set of $k$-symmetric objects; let it also denote the corresponding structure with vocabulary $\{ \in, \emptyset \}$.

Following \cite{blass:apal1999} a {\em $k$-molecule} is an injective mapping $\sigma: k \rightarrow I_k$, i.e. a sequence of $k$ distinct atoms. For a finite sequence of such $k$-molecules $\bar{\sigma} = (\sigma_0, \dots, \sigma_{\ell-1})$ of length $\ell$, the {\em configuration} $\textit{conf\/}(\bar{\sigma})$ is an equivalence relation on $\ell \times k$ defined by $(i,p)\; \sim_{\bar{\sigma}} \; (j,q) \;\Leftrightarrow\; \sigma_i(p) = \sigma_j(q)$. A configuration describes how the $k$-molecules in the sequence $\bar{\sigma}$ overlap. We see that $\textit{conf\/}(\bar{\sigma})$ is uniquely determined by the configurations $\textit{conf\/}(\sigma_i, \sigma_j)$ for $i \neq j$. 

For $\ell \in \mathbb{N}$, $\ell \neq 0$ an {\em abstract $\ell$-configuration} is an eqivalence relation on $\ell \times k$ satisfying $(i,p) \sim (i,q) \Leftrightarrow p = q$. Every configuration $\textit{conf\/}(\bar{\sigma})$ is an abstract $\ell$-configuration. Conversely, given an abstract $\ell$-configuration, choose a different atom $x_{(i,p)}$ for the equivalence class $[(i,p)]_{\sim}$, so $\sigma_i(p) = x_{(i,p)}$ defines a configuration $\bar{\sigma} = (\sigma_0, \dots, \sigma_{\ell-1})$ that realises the abstract $\ell$-configuration.

The set of {\em $k$-forms} is the smallest set $\mathcal{F}$ with (1) $\{ c_0, \dots, c_{k-1} \} \subseteq \mathcal{F}$, where the $c_p$ are new symbols, and (2) whenever $\varphi_1, \dots, \varphi_n \in \mathcal{F}$ and $E_1, \dots, E_n$ are abstract $2$-configurations, then the set of pairs $\varphi = \{ (\varphi_i, E_i) \mid 1 \le i \le n \}$ is a form in $\mathcal{F}$. Each $k$-form $\varphi \in \mathcal{F}$ has a {\em rank} $rk(\varphi)$. We have $rk(c_p) = 0$ and $rk(\{ (\varphi_i, E_i) \mid 1 \le i \le n \}) = 1 + \max \{ rk(\varphi_i) \mid 1 \le i \le n \}$.

A $k$-molecule $\sigma$ together with a $k$-form $\varphi \in \mathcal{F}$ defines a unique object $\varphi * \sigma \in \textit{HF\/}(I)$:

\begin{itemize}
\item For $\varphi = c_p$ we have $\varphi * \sigma = \sigma(p)$;

\item For $\varphi = \{ (\varphi_i, E_i) \mid 1 \le i \le n \}$ we have $\varphi * \sigma = \{ \varphi_i * \tau \mid E_i = \textit{conf\/}(\tau, \sigma) \}$.
\end{itemize}

Then the proofs of Lemmata 36-40 in \cite{blass:apal1999} remain valid without change.

\begin{lemma}\label{lem-form}

For any automorphism $\pi \in Aut(I_k)$ and any $k$-molecule $\sigma$ we have $\pi(\varphi * \sigma) = \varphi * \pi \sigma$.

\end{lemma}

If $\pi$ pointwise fixes $range(\sigma)$, Lemma \ref{lem-form} implies $\pi (\varphi * \sigma) = \varphi * \pi \sigma = \varphi * \sigma$, i.e. $range(\sigma)$ is a support set of $\varphi * \sigma$ of size $\le k$.

\begin{lemma}\label{thm-form}

Every $k$-symmetric object $x \in I_k$ can be written in the form $x = \varphi * \sigma$ with a $k$-form $\varphi$ and a $k$-molecule $\sigma$.

\end{lemma}

\begin{lemma}\label{lem-conf-ext}

Let $m \ge 3$ and assume $| I_k | \ge km$. Let $\bar{\sigma} = \sigma_1, \dots \sigma_\ell$ and $\bar{\tau} = \tau_1, \dots, \tau_\ell$ be sequences of $k$-molecules over $I$ and $J$, respectively, with $\ell < m$. If $\textit{conf\/}(\bar{\sigma}) = \textit{conf\/}(\bar{\tau})$ holds and $\sigma_0$ is another $k$-molecule over $I_k$, then there exists a $k$-molecule $\tau_0$ over $J$ with $\textit{conf\/}(\sigma_0,\bar{\sigma}) = \textit{conf\/}(\tau_0,\bar{\tau})$.

\end{lemma}

Then we can express relationships between elements of $k$-symmetric objects using relations over forms and abstract configurations that do not depend on the input structure. The decisive point is that the $k$-molecules needed to construct the $k$-symmetric objects only enter via their configurations. 

\begin{lemma}\label{lem-relations}
There exist ternary relations $In$ and $Eq$ such that for every structure $I_k$ we have
\begin{align}
\psi * \tau \in \varphi * \sigma \quad &\Leftrightarrow \quad In(\psi,\varphi,\textit{conf\/}(\tau,\sigma)) \label{eq-in} \\
\psi * \tau = \varphi * \sigma \quad &\Leftrightarrow \quad Eq(\psi,\varphi,\textit{conf\/}(\tau,\sigma)) \label{eq-eq} 
\end{align}
for all $k$-forms $\varphi, \psi$ and all $k$-molecules $\sigma, \tau$.

\end{lemma}

\subsection{Equivalence Theorem and Limitations}\label{sec:limits}

With the results above we know that a problem with empty input signature is in CPS, if we can find a formula in FO[PFP] that separates the transitive structures $I_k$ accepted by a CPS computation from those that are not accepted. For sufficiently large input sets no such formula exists. This follows from the following Equivalence Theorem.

\begin{theorem}[Equivalence Theorem]\label{thm-equivalence}
If sets $I$ and $J$ are sufficiently large, then the structures $I_k$ and $J_k$ are $\mathcal{L}^\omega_{\omega\infty}$-equivalent.

\end{theorem} The proof is the same as the proof of Theorem 35 in \cite{blass:apal1999}, which exploits the representation of the $k$-symmetric objects from above. Actually, the proof shows that $I_k$ and $J_k$ are $\mathcal{L}^m_{\omega\infty}$-equivalent for $m \ge 3$. A standard result in Finite Model Theory (see e.g. \cite[Thm.~11.5]{libkin:2004}) states that structures are $\mathcal{L}^m_{\omega\infty}$-equivalent iff the duplicator has a winning strategy for the corresponding pebble game, which is defined as follows:

Both spoiler and duplicator have $m$ pebbles numbered $0, \dots, m-1$. In every move the spoiler chooses a structure (in our case $I_k$ or $J_k$) and places one of its pebbles onto an object of this structure. The duplicator responds (if possible) by placing its own pebble with the same number on an object of the other structure. If the sequences of objects $\bar{x} = x_0, \dots, x_{m-1}$ and $\bar{y} = y_0, \dots, y_{m-1}$ covered by pebbles define a partial isomorphism between the two given structures, the game continues. Otherwise the spoiler wins the game. Thus the duplicator has a winning strategy iff it can be guaranteed that there exists always a response move that lets the game continue forever.

In our case for structures $I_k$ and $J_k$ two sequences $\bar{x}, \bar{y}$ define a {\em partial isomorphism} between $I_k$ and $J_k$ iff $x_i = x_j \Leftrightarrow y_i = y_j$ and $x_i \in x_j \Leftrightarrow y_i \in y_j$ hold for all $0 \le i,j < m$.

Finally, exploit that FO[PFP] can be naturally embedded in $\mathcal{L}^\omega_{\omega\infty}$. For a proof refer to standard textbooks \cite{ebbinghaus:1995} or \cite{libkin:2004} on finite model theory and the remarks in \cite[Sect.~2.3]{blass:apal1999} on a generalisation to infinite structures, which apply in the same way to FO[IFP] and FO[PFP]. Then Theorem \ref{thm-equivalence} implies that Parity is not in CPS.

\begin{corollary}

Parity is not in CPS.

\end{corollary}

\begin{proof}

For any PSPACE ASM $\tilde{M}$ with empty input signature the Support Theorem \ref{thm-support} implies that there exists some $k$ such that the structure $I_k$ contains all active objects for the input set $I$. The Fixed-Point Theorem \ref{thm:pfp} further implies that there exists a formula $\varphi$ in FO[PFP] that holds in $I_k$ iff $\tilde{M}$ accepts $I$. Consider $\varphi$ as a formula in $\mathcal{L}^\omega_{\omega\infty}$. Then the Equivalence Theorem \ref{thm-equivalence} implies that large enough input sets $I, J$ are either both satisfied by $\varphi$ or both not. Hence $\varphi$ cannot separate the input sets $I$ with even cardinality from those with odd cardinality.\qed

\end{proof}

\section{Concluding Remarks}\label{sec:schluss}

In this article we introduced deterministic PSPACE-bounded ASMs, which define a complexity class CPS (choiceless polynomial space). We proved that CPS cannot capture PSPACE; it does not even subsume PTIME. The key for the proof is a characterisation by partial fixed-point formulae over the St\"ark/Nanchen logic for deterministic ASMs and a construction of transitive structures, in which such formulae must hold. 

While CPS subsumes choiceless polynomial time (CPT), we did not yet explore fully the extent of CPS. Indeed, it is open whether on arbitrary input structures CPS can be separated from CPT. However, it is rather straightforward to see that on ordered structures CPS will capture PSPACE. For this we can simply simulate a non-deterministic ASM by a deterministic one, in which each choice rule is replaced by selecting the smallest object in the given order. 

Furthermore, in analogy to \cite[Thm.~21]{blass:apal1999} we can show that the class of problems in CPS is non-negligible. For this consider a signature with a unary predicate symbol $P$ and a binary predicate symbol $<$. Take the class $K$ of structures $A$, in which $P^A$ is a small set in the sense that $|P^A|! < |A|^k$ holds for some constant $k$. Consider any problem in PSPACE that would require a naked set $P^A$ as its input structure.

We can define a PSPACE ASM, which first generates all total orders on $P^A$; in fact, this can be done in polynomial time. We obtain $|P^A|!$ different orders, so we continue running PSPACE ASMs in parallel for all these orders. In this way we can solve the given PSPACE problem (on $P^A$) using PSPACE ASMs that simulate PSPACE Turing machines on the ordered input (in parallel for all generated orders). The parallel ASM is a PSPACE ASM because of our assumption on the size of $P^A$.

Recent work in \cite{GradelS19} also studies the choiceless fragment of a space complexity class, namely choiceless logarithmic space (CLogspace).  Though the problem there is of a somewhat different nature, since an approach based on discarding the time bound in CPT and allowing sets with a transitive closure of logarithmically many objects makes it possible to define sets containing logarithmically many atoms, which admits no straightforward evaluation in LOGSPACE. Nevertheless, they are able to define a choiceless logic which subsumes all previously known logics in LOGSPACE. On the other hand, they also show that the choiceless restriction has considerable impact. Even though their choiceless logic includes counting, they prove that it cannot define all queries in LOGSPACE. Earlier works related to ours are \cite{SuciuP97,BiskupPSB04}, where the authors show that the parity query is not expressible in the polynomial-space fragment of the powerset algebra for nested relations and that it is not expressible in the sparse fragment of the equation algebra either. An open research question is how the expressive power of these query languages relate to CPS. 

Last but not least, a natural research direction would be to explore whether CPS with counting can capture PSPACE. Notice that the analogous question of whether CPT plus counting can capture P has received considerable attention (see e.g.~\cite{DawarRR08} and~\cite{Rossman10}). This is in the case of CPT still an open question.  

\bibliographystyle{abbrv} 
\bibliography{CPS}

\begin{thebibliography}{10}

\bibitem{BiskupPSB04}
J.~Biskup, J.~Paredaens, T.~Schwentick, and J.~V. den Bussche.
\newblock Solving equations in the relational algebra.
\newblock {\em {SIAM} J. Comput.}, 33(5):1052--1066, 2004.

\bibitem{blass:beatcs2007}
A.~Blass and Y.~Gurevich.
\newblock Background of computation.
\newblock {\em Bulletin of the {EATCS}}, 92:82--114, 2007.

\bibitem{blass:apal1999}
A.~Blass, Y.~Gurevich, and S.~Shelah.
\newblock Choiceless polynomial time.
\newblock {\em Annals of Pure and Applied Logic}, 100:141--187, 1999.

\bibitem{boerger:ai2016}
E.~B\"{o}rger and K.-D. Schewe.
\newblock Concurrent {Abstract State Machines}.
\newblock {\em Acta Informatica}, 53(5):469--492, 2016.

\bibitem{boerger:fi2020}
E.~B\"{o}rger and K.-D. Schewe.
\newblock A behavioural theory of recursive algorithms.
\newblock {\em Fundamenta Informaticae}, 177(1):1--37, 2020.
\newblock A preliminary version is available at
  http://arxiv.org/abs/2001.01862.

\bibitem{boerger:2003}
E.~B\"orger and R.~St\"{a}rk.
\newblock {\em {Abstract State Machines}}.
\newblock Springer-Verlag, Berlin Heidelberg New York, 2003.

\bibitem{DawarRR08}
A.~Dawar, D.~Richerby, and B.~Rossman.
\newblock Choiceless polynomial time, counting and the
  {Cai-F{\"{u}}rer-Immerman} graphs.
\newblock {\em Ann. Pure Appl. Log.}, 152(1-3):31--50, 2008.

\bibitem{ebbinghaus:1995}
H.-D. Ebbinghaus and J.~Flum.
\newblock {\em Finite Model Theory}.
\newblock Perspectives in Mathematical Logic. Springer, 1995.

\bibitem{ferrarotti:tcs2016}
F.~Ferrarotti, K.-D. Schewe, L.~Tec, and Q.~Wang.
\newblock A new thesis concerning synchronised parallel computing -- simplified
  parallel {ASM} thesis.
\newblock {\em Theor. Comp. Sci.}, 649:25--53, 2016.

\bibitem{ferrarotti:amai2018}
F.~Ferrarotti, K.-D. Schewe, L.~Tec, and Q.~Wang.
\newblock A unifying logic for non-deterministic, parallel and concurrent
  {Abstract State Machines}.
\newblock {\em Ann. Math. Artif. Intell.}, 83(3-4):321--349, 2018.

\bibitem{GradelS19}
E.~Gr{\"{a}}del and S.~Schalth{\"{o}}fer.
\newblock Choiceless logarithmic space.
\newblock In P.~Rossmanith, P.~Heggernes, and J.~Katoen, editors, {\em 44th
  International Symposium on Mathematical Foundations of Computer Science,
  {MFCS} 2019, August 26-30, 2019, Aachen, Germany}, volume 138 of {\em
  LIPIcs}, pages 31:1--31:15. Schloss Dagstuhl - Leibniz-Zentrum f{\"{u}}r
  Informatik, 2019.

\bibitem{gurevich:cttcs1988}
Y.~Gurevich.
\newblock Logic and the challenge of computer science.
\newblock In E.~B\"orger, editor, {\em Current Trends in Theoretical Computer
  Science}, pages 1--57. Computer Science Press, 1988.

\bibitem{gurevich:tocl2000}
Y.~Gurevich.
\newblock Sequential {Abstract State Machines} capture sequential algorithms.
\newblock {\em ACM Trans. Comp. Logic}, 1(1):77--111, 2000.

\bibitem{libkin:2004}
L.~Libkin.
\newblock {\em Elements of Finite Model Theory}.
\newblock Texts in Theoretical Computer Science. An {EATCS} Series. Springer,
  2004.

\bibitem{Rossman10}
B.~Rossman.
\newblock Choiceless computation and symmetry.
\newblock In A.~Blass, N.~Dershowitz, and W.~Reisig, editors, {\em Fields of
  Logic and Computation, Essays Dedicated to Yuri Gurevich on the Occasion of
  His 70th Birthday}, volume 6300 of {\em Lecture Notes in Computer Science},
  pages 565--580. Springer, 2010.

\bibitem{staerk:jucs2001}
R.~St\"ark and S.~Nanchen.
\newblock A logic for {Abstract State Machines}.
\newblock {\em Journal of Universal Computer Science}, 7(11), 2001.

\bibitem{SuciuP97}
D.~Suciu and J.~Paredaens.
\newblock The complexity of the evaluation of complex algebra expressions.
\newblock {\em J. Comput. Syst. Sci.}, 55(2):322--343, 1997.

\end{thebibliography}

\end{document}